\documentclass{edbk}
\usepackage{edbkps}
\usepackage{epsfig}
\let\footnote\savefootnote
\let\footnoterule\savefootnoterule 
\normallatexbib

\begin{document}

\articletitle{Normal State Properties of Cuprates:\\
$\lowercase{t}$-$J$ Model vs. Experiment }

\author{P. Prelov{\v s}ek$^{1,2}$}
\affil{$^1$J. Stefan Institute, SI-1000 Ljubljana, Slovenia\\
$^2$Faculty of Mathematics and Physics, University of Ljubljana,
SI-1000 Ljubljana, Slovenia}

\begin{abstract}
We discuss some recent results for the properties of doped
antiferromagnets, obtained within the planar $t$-$J$ model mainly by
the finite-temperature Lanczos method, with the emphasis on the
comparison with experimental results in cuprates. Among the
thermodynamic properties the chemical potential and entropy are
considered, as well as their relation to the thermoelectric power. At
the intermediate doping model results for the optical conductivity,
the dynamical spin structure factor and spectral functions reveal a
marginal Fermi-liquid behaviour, close to experimental findings. It is
shown that the universal form of the optical conductivity follows
quite generally from the overdamped character of single-particle
excitations.
\end{abstract}

\section{Introduction}

It is by now quite evident through numerous experiments on electronic
properties that cuprates, being superconductors at high temperatures,
are also strange metals in the normal phase \cite{imad}. On the other
hand it also appears that most features can be well represented by
prototype single-band models of correlated electrons, as the Hubbard
model and the $t$-$J$ model. In spite of their apparent simplicity
these models are notoriously difficult to treat analytically, in
particular in the most interesting regime of strong correlations. This
has led to intensive efforts towards numerical approaches \cite{dago},
mostly using quantum Monte Carlo (QMC) and the exact diagonalization
(ED) methods.

The subject of this contribution is the planar $t$-$J$ model (Hubbard
model is expected to show similar behaviour in the strong correlation
regime), which represents layered cuprates as doped antiferromagnets
(AFM) and within a single band both mobile charges and spin degrees of
freedom,
\begin{equation}
H=-t\sum_{\langle ij\rangle  s}(\tilde{c}^\dagger_{js}\tilde{c}_{is}+
{\rm H.c.})+J\sum_{\langle ij\rangle} ({\bf S}_i\cdot {\bf S}_j -
{1\over 4} n_i n_j). \label{eq1}
\end{equation} 
Strong correlations are here imposed by strictly forbidding doubly
occupied sites. So far most calculations were performed for the ground
state at $T=0$, where the standard Lanczos algorithm offers an
efficient exact-diagonalization analysis of small systems \cite{dago}.
More recently a novel numerical method, finite-temperature Lanczos
method (FTLM) \cite{jplan,jprev}, has been introduced combining the
Lanczos method with a random sampling, which allows for an analogous
treatment of static and dynamic properties at $T>0$.

One of most important conclusion of experimental efforts in the last
decade is the realization that the electronic phase diagram in the
paramater space of planar hole concentration $c_h$ and temperature $T$
is quite universal.  Materials are usually classified as underdoped,
optimally doped and overdoped, with respect to the highest $T_c$ in a
given class. In our analysis we cannot establish the
superconductivity, so we will use the highest entropy (at low $T$) as
a criterion for the optimum doping. In fact both criteria are quite
close for real cuprates \cite{coop} and one could conjecture that this
relation is not accidental. In particular since in thermodynamic
quantities at the same doping also the pseudogap scale disappears.

In this contribution we mainly discuss two topics related to the
normal-state properties of cuprates. In Sec.II we deal with the
thermodynamic quantities: entropy, chemical potential and closely
related thermoelectric power, all of them in the full range of
$c_h$. In Sec.III we concentrate on the appearance and the relation
between different manifestations of the marginal-Fermi-liquid (MFL)
behaviour, observed in the optical conductivity, dynamic spin
susceptibility and spectral functions at the intermediate (optimum)
doping.

\section{Thermodynamics}

Let us first consider thermodynamic quantities, which can be directly
derived from the grand-canonical sum: free energy density ${\cal F}$,
chemical potential $\mu$ and entropy density $s$. For these the FTLM
is particularly simple to implement \cite{jpterm,jprev}, since it
requires only a minor generalization of the usual Lanczos method.
Results presented below were obtained mostly for $N=20$ sites and
parameters $J/t=0.3$, corresponding to the situation in cuprates
(where $t \sim 0.4$eV). Note that in small systems only results above
certain (size dependent) $T$ are meaningful, i.e. in cases below
typically $T>T_{fs} \sim 0.1~t$.

\begin{figure}
\center
{\epsfig{file=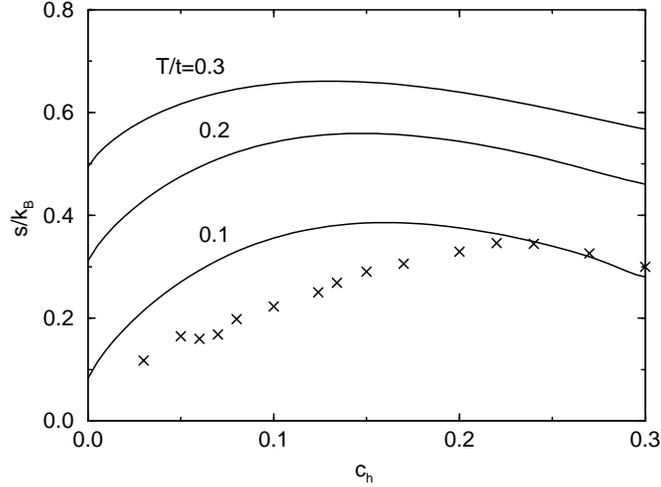,height=10cm,angle=-90}}
\caption{ $s$ vs. $c_h$ at several $T$ \cite{jprev}. For comparison also
experimental results for LSCO \cite{coop} at highest $T=320~K \sim
0.07~t$ are shown.  } \label{fig1}
\end{figure}

Let us first discuss results for $s$, shown in Fig.(\ref{fig1}) for
various $T$. It seems quite generic feature of such a system that
$s(c_h)$ exhibits a (rather broad) maximum at the intermediate doping
$c_h \sim c_h^*$. The increase of $s$ on doping can be plausibly
related to the frustration between the AFM exchange $\propto J$ and
the hole kinetic energy $\propto c_h t$ preferring the FM
configuration. This naturally leads to a most frustrated situation at
$c_h^* \sim J/t$. It is quite fortunate that the FTLM works best,
i.e. $T_{fs}$ is lowest, just in the cases with large $s$ and large
frustration, while other methods have difficulties in such
situation. E.g. QMC is plagued with the minus-sign problem which seems
to be intimately related to fermionic frustration.

Even more surprising fact is the magnitude of $s$ at $T<J$. E.g. at
$T=0.1~t$ at $c_h^*$ we get 40\% of $s(T=\infty)$. Clearly we are
dealing with a system which has very low degeneracy (Fermi)
temperature $T_{deg}<J$, far below the free fermion value $T_{deg}^0
\sim 8t$. Such a conclusion is in agreement with experiments in
cuprates. In recent years $s$ has been measured in YBCO and LSCO (also
presented in Fig.(\ref{fig1}) in a wide doping regime \cite{coop} and
our results show good quantitative agreement.

\begin{figure}
\center
{\epsfig{file=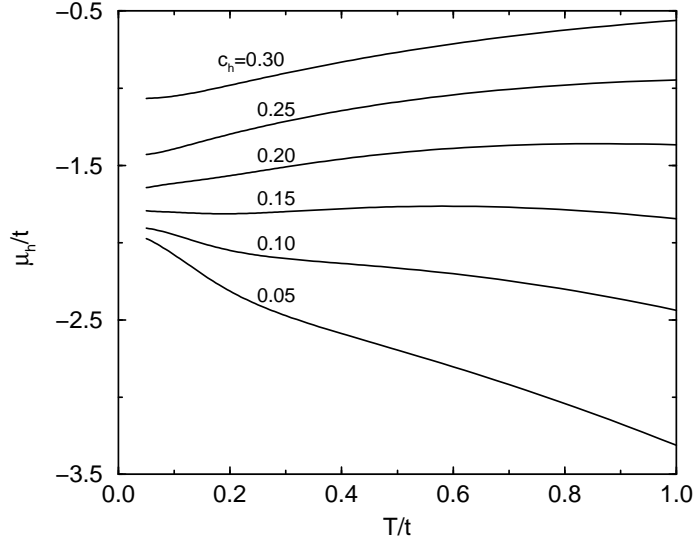,height=10cm,angle=-90}}
\caption{
Hole chemical potential $\mu_h$ vs. $T$ at several dopings $c_h$
\cite{jprev}.  }
\label{fig2}
\end{figure}

For $\mu_h(T)$, presented in Fig.(\ref{fig2}), we mostly do not find a
$T^2$ dependence of $\mu_h$ at low $T$, as expected for a normal Fermi
liquid, except within the extremely overdoped regime $c_h\geq 0.3$. In
particular, in a broad range $0.05 <c_h <0.3$ we find a roughly linear
variation $\mu_h(T)= \mu_h(T=0) + \alpha k_B T$, whereby the slope
$\alpha$ changes the sign at $c_h = c_h^* \sim 0.15$.  The variation 
$\mu(c_h)$ at low $T$ has been recently deduced
experimentally from the shift of photoemission spectroscopy spectra in
LSCO \cite{ino}, and the agreement with our results is quite
satisfactory \cite{jprev}. From photoemission results as well as from
our Fig.(\ref{fig2}) it is also evident that $\mu(c_h<c_h^*)$ is very
flat which would indicate that at $T \to 0$ and $c_h<c_h^*$ the
compressibility $\kappa \propto -dc_h/d\mu$ is very large or even
diverging, as would e.g. follow from the phase separation scenario
\cite{emer} or the singular $c_h \to 0$ limit \cite{imad}. It should
be however stressed that the distinction between these scenarios could
be relevant only at very low $T\ll J$, since both experiment and
numerics indicate on quite large $s$, i.e. a distribution over a wide
spectrum of states, even at $T \sim J/10$.

It is quite helpful to realize that the free energy density ${\cal
F}(c_h,T)$ relates the variation of $s=\partial{\cal F}/\partial T$
and $\mu=\partial{\cal F}/\partial c_h$, i.e.
\begin{equation}
\left. {\partial s\over \partial c_h}\right|_T=\left. -
{\partial \mu_h\over \partial T}\right|_{c_h}=
\frac{\partial^2{\cal F}}{\partial c_h \partial T} . \label{eq2}
\end{equation}
This connects the maximum $s(c_h^*)$ with the change in slope
$d\mu_h(c_h^*)/dT=0$.  Moreover Eq.(\ref{eq2}) allows us to discuss
more confidently the slope $d\mu_h/dT= \alpha k_B$ for which we find
in the underdoped regime $\alpha \sim 2$. Although the latter has not
been so far verified directly for cuprates, one can extract in the
same regime from the measured $\partial s/ \partial c_h$ for LSCO and
YBCO at $T>100~K$ similar values $\alpha >1$ \cite{coop}. It is quite
evident that at $\alpha>1$ we are not dealing with a degenerate Fermi
liquid but rather with the nondegenerate doped carriers, which is a
situation typical for a doped (nondegenerate) semiconductor. One
should just recall the standard expression for $\mu_h$ in p-type
semiconductor,
\begin{equation}
c_h=P_v {\rm e}^{-(\epsilon_v-\mu_h)/k_BT} \Longrightarrow
-\frac {\partial \mu_h}{\partial T}= k_B {\rm ln}\frac{P_v}{c_h} 
>k_B, \label{eq3}
\end{equation} 
where in our notation $P_v \sim 1$. The constant slope $d\mu_h/dT$
observed in our calculations down to $T < 0.1~t$ is a confirmation of
such a picture. In experimental results for $s$ one should however
notice a reduction of $\alpha$ with $T$, but even at $T\sim T_c$ the
system is not evidently a normal Fermi liquid with $\alpha <1$.

\begin{figure}
\center
{\epsfig{file=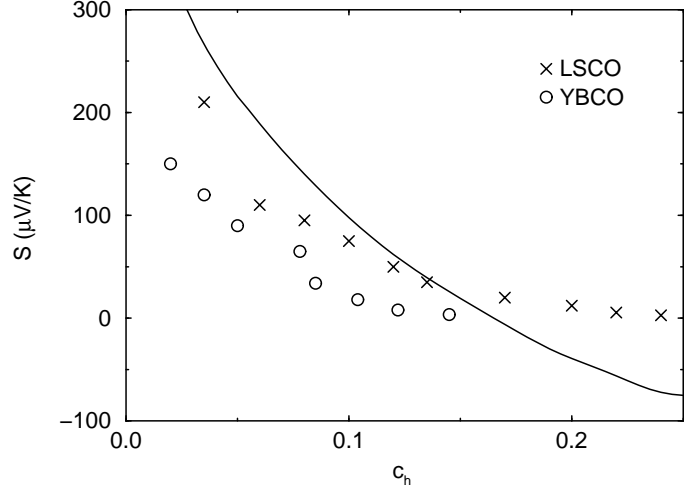,height=10cm,angle=-90}}
\caption{ Thermoelectric power $S$ vs. $c_h$ for $T/t=0.1$ \cite{jprev}.
Experimental result for LSCO and oxygen deficient YBCO are taken from 
Ref.~\cite{coop}.  } \label{fig3}
\end{figure}

Another consequence of such a semiconductor picture is an expression
for the thermopower $S$,
\begin{equation}
S \sim \frac {\bar\epsilon-\mu_h(T)}{ e_0 T} \sim
\frac {\mu_h(T=0)-\mu_h(T)}{ e_0 T} \sim \alpha S_0,\label{eq4}
\end{equation}
where $S_0= k_B/e_0 = 86 \mu V/K$. The validity of this approximation
we have verified within the $t$-$J$ model also directly by evaluating
the mixed current-energy current correlation function and observing
that they are proportional to the current-current correlation
$C_{j_Ej}(\omega) = \mu_h(0) C(\omega)$. The result is in
Fig.(\ref{fig3}) is good agreement with the general experimental
observation in cuprates \cite{kais} of a large and rather
$T$-independent $S$ at low doping. In fact instead of the usual
semiconductor expression for $\alpha$ at $c_h \ll 1$, Eq.(\ref{eq4}),
in a strongly correlated system it is more appropriate to use the proper
statistics for the $t$-$J$ model leading to $\alpha \sim {\rm
ln}[2(1-c_h)/c_h]$, which even predicts the change of sign at $c_h
\sim 0.3$.

\section{Dynamics at optimum doping}

It has been quite early established from experiments that cuprates
in the normal state do not follow the behaviour consistent with the
normal Fermi liquid. In contrast several static and dynamic quantities
at optimum doping can be quite well accounted for within the marginal
Fermi liquid (MFL) concept \cite{varm}. Most evident example is the
dynamic conductivity $\sigma(\omega)$ which does not obey the usual
Drude form with a constant rate $1/\tau$ but can be well fitted in a
broad range of $\omega,T$ with the generalized MFL form
$\tau^{-1}(\omega,T)= \tilde \lambda( |\omega| + \eta T)$, describing
also the well established linear resistivity law $\rho \propto T$.  It
has been natural to postulate an analogous MFL behaviour for
quasiparticle (QP) relaxation in spectral functions as e.g. measured
by the angle-resolved photoemission spectroscopy (ARPES). Only
recently, however, the high resolution ARPES experiments on BSCCO
\cite{valla} seem to be in position to confirm beyond doubt this
behaviour, obeyed in the optimum-doped materials surprisingly even at
$T <T_c$ for QP along the nodal direction in the Brillouin zone.  Most
evident indication that also spin dynamics follows the MFL concept is
the observed anomalous NMR and NQR spin-lattice relaxation rate
$1/T_1(T) \sim const.$ \cite{imai} instead of usual Korringa law in
metals.

By calculating using FTLM several related quantities, descibing charge
and spin dynamics within the $t$-$J$ model, we established that the
MFL concept applies well in a broad range of intermediate hole doping
$0.1<c_h<0.3$.  We discuss here in particular the dynamical
conductivity $\sigma(\omega)$, the local spin susceptibility
$\chi_L(\omega)$ \cite{jpuni} and the QP relaxation rate as obtained
from the analysis of spectral functions $A({\bf k},\omega)$
\cite{jpspec}.  Moreover, $\sigma(\omega)$ has been been found close
to a universal form \cite{jpuni},
\begin{equation}
\sigma(\omega) =  C_0\frac{1 - e^{-\omega/k_B T}}{\omega},
\label{eq5}
\end{equation}
in a remarkably broad frequency regime $0<\omega < \omega^* \sim 2t$,
while $C_0$ is essentially $T$-independent for $T<J$.  Resulting
$\sigma(\omega<\omega^*)$ is clearly governed by $T$ only.
Evidently, Eq.(\ref{eq5}) reproduces the linear resistivity law
$\rho=T/C_0$ and is consistent with the MFL scenario for
$\tau^{-1}(\omega,T)$, however in a very restrictive way since both
MFL parameters are essentially fixed. A reasonable overall fit can be
e.g. achieved by $\tilde \lambda \sim 0.6$ and $\eta \sim 2.7$.  When
optical experiments on $\sigma(\omega)$ in cuprates are analysed
within the MFL framework quite close values for $\tilde \lambda, \eta$
are in fact reported \cite{schl,elaz}. In addition, the model results
reproduce well also the absolute value of $\sigma(\omega)$ and
$\rho(T)$ \cite{jprev}.

Analogous universality has been found also in the spin dynamics, in
particular when looking at the local spin susceptibility
$\chi_L(\omega)$ and related spin correlation function $S_L(\omega)$,
\begin{equation}
\chi''_L(\omega) = {1\over \pi}\tanh\left( {\omega \over 2T} 
\right) \bar S_L(\omega), \label{eq6}
\end{equation}
where $\bar S_L(\omega)=S_L(\omega)+S_L(-\omega)$ is the symmetrized
function, having a fixed sum rule
\begin{equation}
\int_0^{\infty}  \bar S(\omega) d\omega = \langle (S_i^z)^2\rangle = 
{1 \over 4} (1-c_h). \label{eq7}
\end{equation}
The most important message of numerical results on spin dynamics at
intermediate doping is that $\bar S_L(\omega)$ is nearly
$T$-independent in a broad range of $T$, in particular for $T<J$.
Moreover $\bar S_L(\omega)$ is only weakly doping dependent consistent
with the sum rule. So we have a conclusion that at intermediate doping
the more fundamental and universal quantity is the correlation
function $S_L(\omega)$ and not the susceptibility $\chi_L(\omega)$,
which is the analogy to the relation between $C(\omega)$ and
$\sigma(\omega)$ in Eq.(\ref{eq5}). For the spin dynamics this is also
the message of the anomalous NMR $1/T_1$ in cuprates \cite{imai}. As a
result $\chi_L''(\omega)$, Eq.(\ref{eq6}), follows the MFL behaviour
i.e. at $\omega<T$ one gets anomalous $T$ dependence $\chi_L''(\omega)
\propto \omega/T$.

A MFL-type QP relaxation is extracted within the $t$-$J$
model also from the analysis of the spectral functions $A({\bf k},\omega)$
near the optimum doping \cite{jpspec,jprev}. For the characterization
of QP properties the self energy $\Sigma({\bf k},\omega)$ is crucial.
On the other hand the same information can be also expressed in terms
of QP parameters $Z_{\bf k},\Gamma_{\bf k},\epsilon_{\bf k}$. Both
definitions are related as
\begin{equation}
A({\bf k}, \omega)=-\frac{1}{\pi}{\rm Im} \frac{1}{\omega -
\Sigma({\bf k},\omega)} =\frac{1}{\pi}\frac{Z_{\bf k}\Gamma_{\bf k}}
{(\omega-\epsilon_{\bf k})^2+\Gamma^2_{\bf k}}. \label{eq8}
\end{equation}
For QP near the Fermi surface the hole-part self energy $\omega<0$ is
found to be of the MFL form, i.e. Im$\Sigma \sim -\tilde
\gamma(\omega+\xi T)$ with $\tilde \gamma \sim 1.4$ and $\xi \sim
3.5$.  $\tilde \gamma > 1$ means an overdamped character of QP, since
the the full width at half maximum $\Delta \sim
2\Gamma(\epsilon)>\epsilon$ is larger than the QP (binding) energy
$\epsilon$.  This should be contrasted with the electron-like regime
$\omega>0$ where the damping is found to be essentially smaller and
consequently QP can be underdamped.

Here we comment on the relation of our results to recent ARPES results
in BSCCO. The analysis for hole-like excitations in the nodal
direction $(0,0) - (\pi,\pi)$ shows the MFL form with the QP damping
$\Gamma \sim 0.75 \omega$ for $\omega>T$ and $\Gamma \sim 2.5 T$ for
$\omega<T$ \cite{valla}. This again means an overdamped character of
hole excitations, since $2\Gamma(\epsilon)>\epsilon$. In making
the comparison one should take into account that $\Gamma = Z |{\rm
Im}\Sigma|$. Since at the peak position we find $Z \sim 0.5$
experimental and model values appear reasonably close.

Let us finally discuss the relation of $\sigma(\omega)$ and the
associated relaxation rate $1/\tau$ to the QP damping $\Gamma$
\cite{prel}. In the case of weak scattering one finds $1/\tau \sim 2
\Gamma$. In cuprates as well in the $t$-$J$ model we are apparently
dealing with overdamped QP, so the relation is at least
questionable. Also, the conductivity form Eq.(\ref{eq5}) appears to be
universal, while the QP damping does not seem to be parameter free.

\begin{figure}
\center{\epsfig{file=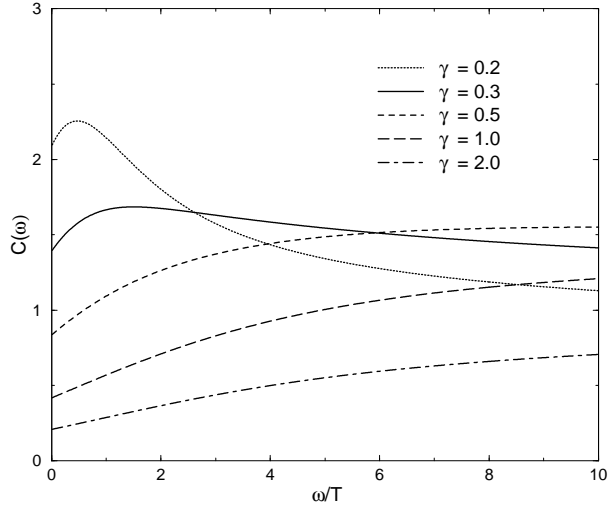,height=8cm,angle=-90}}
\caption{ Current-current correlation spectra $C(\omega)$
vs. $\omega/T$ for various $\gamma$ at fixed $\xi= \pi$.}\label{fig4}
\end{figure}

One approach is to approximate the current-current correlation
function $C(\omega)$, which in general replaces $C_0$ in
Eq.(\ref{eq5}), by a decoupling in terms of spectral functions $A({\bf
k}, \omega)$ neglecting possible vertex corrections, i.e.
\begin{equation}
C(\omega)= \frac{2\pi e_0^2}{N} \sum_{\bf k} (v_{\bf k}^\alpha)^2 \int
{d\omega^\prime} f(-\omega^\prime) f(\omega^\prime-\omega) 
A({\bf k},\omega^\prime) A({\bf k},\omega^\prime-\omega).
\label{eq9}
\end{equation}
In order to reproduce the MFL form of $\sigma(\omega)$ one has to
assume the MFL form for the spectral function, Eq.(\ref{eq8}),
i.e. the QP damping of the form $\Gamma = \gamma (|\omega| + \xi T)$.
We neglect also the ${\bf k}$ dependence of $\Gamma$ and $Z$.  In fact
it is enough to assume that $\Gamma_{\bf k}(\omega)$ is independent of
deviations $\Delta {\bf k}_{\perp}$ perpendicular to the Fermi
surface. The latter is just what is observed in recent ARPES studies
of BSCCO \cite{valla}.  Replacing in Eq.(\ref{eq9}) the ${\bf k}$
summation with an integral over $\epsilon$ with a slowly varying
density of states we can derive
\begin{equation}
\int d\epsilon A(\epsilon,\omega') A(\epsilon,\omega'-\omega)=
\frac{Z^2}{\pi} \frac {\bar \Gamma(\omega,\omega')}
{\omega^2+ \bar \Gamma(\omega,\omega')^2} , \label{eq10}
\end{equation}
where $\bar \Gamma(\omega,\omega')=\Gamma(\omega')
+\Gamma(\omega'-\omega)$.  We are thus dealing with a function
$C(\omega)$ depending only on the ratio $\omega/T$, and on MFL
parameters $\gamma,\xi$. For $\gamma \ll 1$ we recover via such an
analysis $C(\omega)$ strongly peaked at $\omega = 0$ and consequently
MFL-type $\sigma(\omega)$ with $1/\tau(\omega)=2\Gamma(\omega/2)$
\cite{varm}.  No such simple relation is valid when one approaches the
regime of overdamped QP excitations $\gamma \sim 1$ or more
appropriate $\gamma\xi \sim 1$. In Fig.(\ref{fig4}) we show results
for several $\gamma$ fixing $\xi=\pi$. For $\gamma < 0.2$ still a
pronounced peak shows up at $\omega \sim 0$, on the other hand
$C(\omega)$ becomes for $\gamma > 0.3$ nearly constant or very slowly
varying in a broad range of $\omega/T$.

The main message of the above simple analysis is that for systems with
overdamped QP excitations the universal form (\ref{eq5}) describes
quite well $\sigma(\omega)$ for a wide range of parameters. It should
be stressed that nearly constant $C(\omega<\omega^*)$ also means that
the current relaxation rate $1/\tau^*$ is very large, $1/\tau^* \sim
\omega^* \gg 1/\tau$, i.e. much larger than the conductivity
relaxation scale apparent from Eq.(\ref{eq5}) where $1/\tau \propto
T$ follows solely from thermodynamics.

One should also be aware of the upper cutoff scale $\omega^*$ for the
validity of the MFL-like QP damping. In the problem considered here it
appears that the cutoff is directly related to the current relaxation
rate $\omega^* \sim 1/\tau^*$ found in the $t$-$J$ model to be
extremely high at the intermediate doping, i.e. $\omega^* \sim
2t$. The latter allows for an effective mean free path $l^*$ of only
few cells, essentially independent of $T$. Such a short $l^*$ can be
plausibly explained by assuming that charge carriers - holes entirely
loose the phase coherence in collisions with each other due to the
randomizing effect of an incoherent spin background.  Note again that
the short correlation length (even at $T < T_c$) appears also from the
analysis of ARPES spectral functions $A({\bf k},\omega)$ varying
$\Delta {\bf k}_\parallel$ along the Fermi surface \cite{valla}.

\section{Discussion}

Cuprates in their metallic phase are anomalous in several respects.
One important conclusion at least for theoreticians is that most of
anomalous properties are quite well reproduced also in the prototype
$t$-$J$ model. The analysis of this model has been so far restricted
to numerical calculations of small systems, nevertheless in the
$T>T_{fs}$ window where macroscopic relevance of FTLM results is
expected the agreement with experiments is even quantitative, without
any adjustable parameters. Since the behaviour found experimentally is
quite generic and universal down to lowest $T\sim T_c$ there is no
reason to doubt in the generality of model results.

Nevertheless there are open questions of the existence and the origin
of low energy scales in cuprates as well as in the $t$-$J$ model. In
the underdoped or weakly doped regime FTLM shows the indication for
the pseudogap scale $T^*$, in particular in the uniform susceptibility
$\chi_0$ and in the density of states \cite{jprev}. This scale seems
to be related to the onset of short range AFM correlations, hence $T^*
\propto J$. Still for $T<T^*$ the entropy remains large as manifested
by experiments and our results. The electron liquid is thus closer to
a nondegenerate system of composite particles than to a degenerate
Fermi gas. Only at $T\to 0$ the entropy is low enough to make the
discussion of possible orderings or instabilities relevant.

The origin of the MFL behaviour of several dynamic quantities and of
the universal form of $\sigma(\omega)$ and $\chi_L(\omega)$ in the
intermediate doping has to be intimately related to the large
degeneracy in this regime. It has been shown that QP are essentially
overdamped down to lowest $T\sim T_c$. This can be only explained by
the scattering on spin fluctuations, which mainly contribute to the
entropy. On the other hand spins are just strongly perturbed by holes
introduced by doping, so a self-consistent enhancement seems to be the
mechanism which dominates the relevant physics. Only at low $T \sim
T_c$ apparently this behaviour breaks down by an emergence of 
coherence and new energy scales.

\end{document}